\documentclass[%
 reprint,
amsmath,amssymb,
 aps,
]{revtex4-1}

\usepackage{graphicx}
\usepackage{dcolumn}
\usepackage{bm}
\usepackage{epsfig,graphics,amsmath,amssymb}
\usepackage{epstopdf}
\usepackage{upgreek}
\usepackage{booktabs}
\usepackage{color}
\usepackage{draftwatermark}

\begin{document}
\preprint{APS/123-QED}

\title{How to obtain complex transition dipole moments satisfying crystal symmetry and periodicity from \emph{ab-initio} calculations}

\author{Shicheng Jiang$^{1,2}$, Chao Yu$^{1}$, Jigen Chen$^{3}$, Yanwei Huang$^{2}$, Ruifeng Lu$^{1,\dagger}$, and C. D. Lin$^{4,\ast}$}

\affiliation{$^{1}$ Department of Applied Physics, Nanjing University of Science and Technology, Nanjing 210094, P. R. China}
\affiliation{$^{2}$ State Key Laboratory of Precision Spectroscopy, East China Normal University, Shanghai 200062, China}
\affiliation{$^{3}$ Zhejiang Provincial Key Laboratory for Cutting Tools ,Taizhou University, Taizhou 31800, China }
\affiliation{$^{4}$ J. R. Macdonald Laboratory, Department of Physics, Kansas State University, Manhattan, Kansas 66506, USA}

\email[Email:]{$^{\ast}$ cdlin@phys.ksu.edu}
\email[Email:]{$^{\dagger}$ rflu@njust.edu.cn}

\date{\today}

\begin{abstract}
Transition dipole moments (TDM) between energy bands of solids deserve special attention nowadays as intense lasers can easily drive non-adiabatic transitions of excited electron wave packets across the Brillouin zones. The TDM is required to be continuous, satisfying crystal symmetry, and periodic at zone boundaries. While present day \emph{ab-initio} algorithms are powerful in calculating band structures of solids, they all introduced random phases into the eigenfunctions at each crystal momentum k. In this paper, we show how to choose a 
 ``smooth-periodic'' gauge where TDMs can be smooth versus k, preserving crystal symmetry, as well as maintaining periodic at zone boundaries. Based on band structure and TDMs in the ``smooth-periodic'' gauge calculated from \emph{ab-initio} algorithms, we revisit
    high-order harmonic generation from MgO which exhibits inversion symmetry and ZnO which has broken symmetry. The symmetry properties of TDMs with respect to k ensure the absence of even-order harmonics in system with inversion symmetry, while the TDM in the `smooth-periodic'' gauge for ZnO is shown to enhance even harmonics that were underestimated in previous simulations. These results reveal the importance of correctly treating the complex TDMs in nonlinear laser-solid interactions which  has been elusive so far. 
\end{abstract}

\maketitle
\section{Introduction}
In quantum mechanics, band theory is the foundation for understanding the structure of solids and their interactions with lights. To interpret a plethora of experiments on all kinds of condensed materials, \emph{ab-initio} computer codes have been developed and the values of such packages are well recognized. In recent years, with the advent of intense lasers and their nonlinear interactions with solids, such ultrafast phenomena like high-order harmonic generation (HHG)\cite{shambhu_nature-phys2011,chini_apl,Vampa_nature,T.T.Lu_nature2015,Hohenleutne_nature2015}, laser-induced charge transfer\cite{Schiffrin, current_Yakovlev,Paasch, Tong}, Bloch oscillation\cite{schubert_naturephoton2014, LiangLi, L. Liu}, laser-controlled dielectrics \cite{Schultze,Schultze_sience}, and ultrafast renormalization \cite{renormal_1,renormal_2,renormal_3} have been widely investigated.
 The equations of motion based on band theory and crystal-momentum representation,
 for example, semiconductor Bloch equations (SBEs)\cite{Golde, Luu_prb, Koch's book} and other extended forms\cite{Vampa_prb,Dirac_Bloch}, have already been used to interpret these nonlinear ultrafast phenomena. However, true quantitative comparison between theoretical results with experiments remains a challenge, despite that the band structure and transition dipoles were calculated from advanced \emph{ab initio} codes.

 In the interaction of strong lasers with solids, non-adiabatic transitions of electrons between bands are important. The excited electron wave packet can even go across the first Brillouin zone. When the carrier is moving along a path in the k-space, the wavepacket will acquire a dynamical phase and a geometry phase\cite{berry, yuguiyao}. Previously, geometric phase was mostly considered only for closed paths for ensuring the theory is gauge-independent. However, as shown recently\cite{HCDu,Yue},  in the SBEs method for high-order harmonic generation in solids, gauge invariance is achieved when the correct phase of the dipole transition moment (TDM) is included in the SBEs. The combination of geometry phase and dipole phase is well-defined whether the system is open or closed\cite{Yue}. Both of these two phases will be encoded in the macroscopic polarization, and thus the photonic signal\cite{Shambhu_OL}.

In order to describe the interaction of strong laser fields with solids in a finite k space   including the phase accumulated by the moving wavepacket, it requires that: (1) the TDMs should be calculated accurately for the whole first Brilloin zone; (2) the TDMs with k-dependent phase should be continuous and periodic with respect to k.  Using \emph{ab-initio} software to calculate  accurate absolute values of TDMs,  Yu et al.\cite{chao} were able to obtain  improved HHG spectra.  However,  since all \emph{ab-initio} algorithms calculated the eigenfunctions at each k separately, random phases are introduced at each k point and the phase of the TDM is not smooth and continuous. By noting the importance of the TDM phase,  Jiang et al.\cite{jiang_prl} obtained the smooth phase analytically using the tight-binding model. They were able to reproduce the orientation dependence of the HHG spectra from ZnO first reported by  S. Ghimire et al. \cite{shambhu_nature-phys2011} and S. Gholam-Mirzaei et al.\cite{chini_apl}. However, the tight-binding model is too primitive and there remains quantitative discrepancies with experiments. It is desirable to obtain ambiguity-free phase for the TDM calculated from \emph{ab-initio} algorithms.

The problem of random phase in TDM has been well recognized and studied for a long time\cite{Sipe, Gonze, Souza,Hjelm} and solutions have been suggested. However, as pointed out by Yakovlev\cite{Yakovlev}, they either do not ensure the periodicity with respect to k or require the evaluation of the so-called covariant derivatives. Recently, a ``smooth procedure'' suggested in Ref. \cite{Hjelm} has been used in harmonic generation in solids\cite{MXWu,CDLiu,HCDu}.  Here, we will show that this widely used ``smooth procedure'' is not robust. The  method will introduce Zak's phase\cite{Zak} into the eigenfunctions which would then break the periodicity of the TDMs.

In this paper we will obtain smooth TDM phase that also satisfies periodic conditions of the crystal by introducing what we will called the `smooth-periodic'' gauge to distinguish it from the ``periodic gauge''. We first summarize in Section II the "smooth procedure". We then address systems with (Section III) and without (Section IV)  inversion symmetry. In this `smooth-periodic'' gauge, the symmetry properties of k-dependent eigenfunctions and TDMs for system with inversion symmetry are retained. Such symmetry properties would ensure the absence of even-order harmonics driven by long pulses even if multi-band excitations are included. Using such new TDMs, we calculated the harmonic spectra of MgO and found improved agreement with earlier experiment. Similarly, we also revisited the HHG spectra of ZnO in the direction with broken symmetry. While even harmonics of ZnO were predicted in
Jiang et al.\cite{jiang_prl}, the signals were much too weak relative to the odd harmonics. With TDMs calculated using the "smooth-periodic'' gauge, the even harmonics were greatly enhanced and become comparable to odd harmonics. We comment that previously even harmonics were modeled in terms of Berry Curvature. With the "smooth-periodic'' gauge, the latter is a consequence of treating correctly the TDM that satisfies smooth and periodic conditions of crystals. A short conclusion is given in Section V to end this paper.

\section{The commonly used ``smooth procedure'' and its deficiency}
 The commonly used ``smooth procedure'' was described in details \cite{Hjelm} by Hjelm and coworkers. It is necessary to show this method briefly here. The Bloch wavefunction is expressed as $\Phi_{m}(k,r)=e^{ikr}u_{m}(k, r)$ where $u_{m}(k,r)$ is periodic  $u_{m}(k,r)=u_{m}(k,r+R)$, with R being the lattice spacing. In most of the \emph{ab-initio} softwares, the periodic part is expanded by plane waves $u_{m}(k,r)=\sum_{h}a(k+G_h)e^{iG_hr}$. In principle, the Bloch functions are periodic in k space $\Phi_{m}(k,r)=\Phi_{m}(k+G,r)$. Here, G is the reciprocal lattice vector. Since the eigenfunctions are obtained separately for different k points which leads to random phases $\varphi_{m}(k)$, $u_{m}^{\prime}(k, r)=u_{m}(k, r) e^{i \varphi_{m}(k)}$. Note that $\varphi_{m}(k)$ is randomly generated and is discrete with respect to $k$.

 In the ``smooth procedure'', a complex number $z_{m}(k)$ is defined by
\begin{equation}
z_{m}(k)=\left|z_{m}(k)\right| \mathrm{e}^{i \alpha_{m}(k)}=\left\langle u_{m}^{\prime}(k, r) |u_{m}^{\prime}(k+\Delta k, r)\right\rangle.
\end{equation}
A new wavefunction is constructed by
\begin{equation}
u_{m}^{\prime\prime}(k+\Delta k, r)=u_{m}^{\prime}(k+\Delta k, r) e^{-i \alpha_{m}(k)}.
\end{equation}
By renaming the function $u_{m}^{\prime\prime}(k+\Delta k, r)$ according to
\begin{equation}
 u_{m}^{\prime\prime}(k+\Delta k, r) \rightarrow u_{m}^{\prime}(k+\Delta k, r),
\end{equation}
the same procedure can be repeated to the next point $u_{m}^{\prime}(k+2\Delta \mathbf{k}, r)$. When this procedure goes over the first Brillouin zone, the phase-modified wavefunction  become continuous with respect to k. We rename the final wavefunction after the ``smooth procedure'' as $u_{m}^{s}(k, r)$ to distinguish it from the original one generated by \emph{ab-initio} software.

 Since $\Delta k$ is small,
\begin{eqnarray}
&\left\langle u_{m}^{\prime}(k, r) |u_{m}^{\prime}(k+\Delta k, r)\right\rangle \nonumber\\
 &\approx e^{\Delta k\left\langle u_{m}(k,r) | \nabla_{k} u_{m}(k,r)\right\rangle+i\left(\varphi_{m}(k+\Delta k)-\varphi_{m}(k)\right)}
\end{eqnarray}
which leads to
\begin{equation}
u_{m}^{\prime \prime}( k+\Delta k,r)=u_{m} (k+\Delta k,r) e^{-\Delta k\left\langle u_{m}(k,r) | \nabla_k u_{m} (k,r)\right\rangle+ i \varphi_{m}(k)}
\end{equation}
As the procedure of Eq. (1-3) goes through the path $k_0\rightarrow k_0+k$, the phase-modified wavefunction becomes
\begin{equation}
u_{m}^{\prime}(k,r)=u_{m}(k,r) e^{i\int_{k_{0}}^{k} d \kappa D_{m m}(\kappa)}e^{i \varphi_{m}\left(k_{0}\right)}
\end{equation}
where $D_{mm}(k)= i\left\langle u_{m}(k,r) | \nabla_k u_{m} (k,r)\right\rangle$ is the Berry connection.

To conclude, this method  forces the wavefunction to be continuous with respect to k, and the new Berry connection $D_{mm}^s(k)= i\left\langle u_{m}^s(k,r) | \nabla_k u_{m}^s (k,r)\right\rangle=0$. Meanwhile, at the same time this method introduces a phase  $\Theta_m(k)=\int_{k_0}^{k} d \kappa D_{m m}(\kappa)+\varphi_{m}\left(k_{0}\right)$ to the eigenfunction. The additional phase $\Theta_m(k)$ will break the periodicity of the eigenfunction. In the followed two sections, we will provide different methods to deal with the non-periodicity for systems with and without inversion symmetry.

\section{Systems with inversion symmetry}
The eigenfunction $u_{m}(k,r)=\sum_{h}a(k+G_h)e^{iG_hr}$ is defined in the ``periodic gauge''\cite{Resta}. As shown by Zak\cite{Zak}, the Zak's phase $\gamma=\int_{k}^{k+G} d \kappa D_{m m}(\kappa)$ is equal  to  zero or $\pi$ in  the``periodic gauge''.  Thus,  for system with inversion symmetry, the simplest way  for the ``smooth procedure'' is to extend it to the second Brillouin zone. In this way, phase difference between the starting point $k_0$ and $k_0+2G$ is zero or $2\pi$, which means that the periodicity of eigenfunctions in k space is 2G.

In this section, rock-salt MgO with inversion symmetry is taken as example to explain our method. Fig. 1(b) shows the band structure of MgO along  (-1,0,-1)$\rightarrow \Gamma$(0,0,0)$\rightarrow $(1,0,1). Figs. 1(c)-(h) are the corresponding TDMs calculated by the ``smooth procedure'' between different pairs of bands. The eigenfunctions are calculated by DFT package in VASP\cite{vasp} using the Perdew-Burke-Ernzeroff GGA functional. The cutoff energy of plane wave is 500 eV. The k-point grids of 20$\times$20$\times$20 with none-zero weight in the first Brillouin zone and 400 points with zero weight along the one dimensional path are used. Since the DFT simulation underestimates the band gap, the conduction bands are shifted to get better agreement with the experimental gap 7.8 eV.  As expected, both the energy bands and TDMs are periodic with 2G.

In Fig. 2, we present the HHG spectra calculated by semiconductor Bloch equations (SBEs) with one valence band (band 2) and two conduction bands (band 3 and band 4) included. The spectra show two plateaus, with the right side of the dashed line being from the recombination of electron-hole pair from the second conduction band (band 4) to the valence band (band 2)\cite{Shambhu_OL}. To the left, which is due to recombination from band 3, the green arrow marks a minimum which is similar to the Cooper minimum in atoms. Such minimum originates from the minimum of the absolute value of dipole moment  between band 2 and band 3. This minimum has also been found  in the TD-DFT simulation in Ref. \cite{Rubio}. Thus it would be of interest to see if this minimum can be observed in experiments if the detected photon energy range can be extended\cite{Shambhu_OL,MgO-NP, MgO-OL}. This kind of minimum will be discussed in details in another paper\cite{jigen}.

\begin{figure}[!htbp]
    \includegraphics[width=3.5in,angle=0]{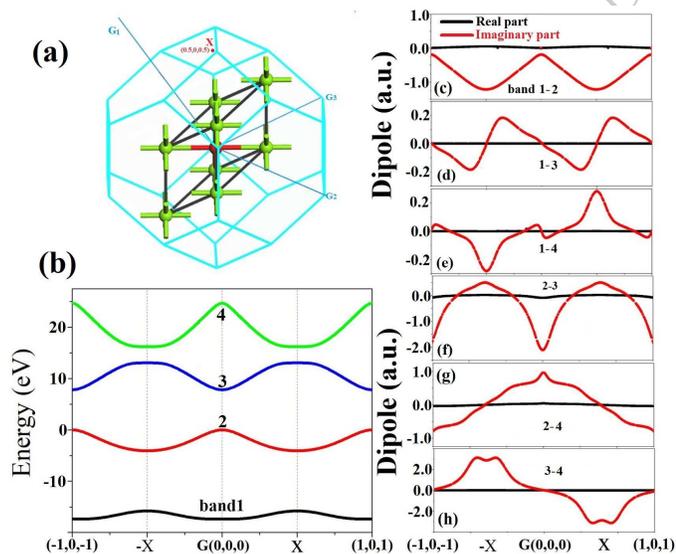}
    \caption{(a) Structure of rock-salt MgO. Red and green balls are O and Mg respectively; (b) Band structures along the $\Gamma-$X axis; (c)-(e) Real (black line) and imaginary (red line) part of TDMs  generated by  the ``smooth procedure'' in the extended Brillouin zone.}\label{Fig_1}
\end{figure}

\begin{figure}[!htbp]
    \includegraphics[width=3.5in,angle=0]{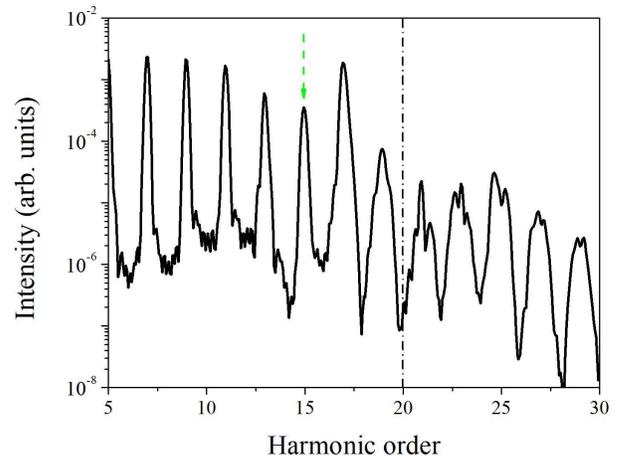}
    \caption{HHG spectra from MgO calculated by 1D three-band SBEs. The equations are solved in the extended Brillouin zone and all the elements used in the SBEs model are from Fig. 1. Laser parameter: 30 fs, 1300nm, 1$\times$10$^{13}$ W/cm$^2$.}\label{Fig_2}
\end{figure}

\begin{figure}[!htbp]
    \includegraphics[width=3.5in,angle=0]{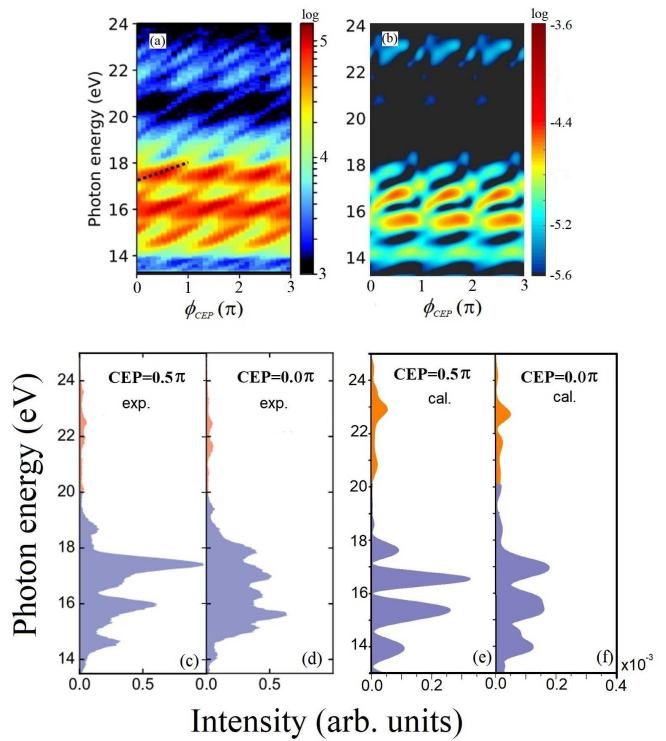}
    \caption{Comparison between experimental and calculated CEP-dependent HHG from MgO. (a) is reprinted from Fig.2 (c) of \cite{Shambhu_OL}. (b) is the calculated spectrum. (c) and (d) are reprinted from Fig. 1(a) of \cite{Shambhu_OL}. (e) and (f) are the calculated spectra along $\Gamma$-X (cut from (b)). Laser parameters: 1700 nm, 10fs, 4$\times$10$^{13}$ W/cm$^2$.}\label{Fig_3}
\end{figure}

\begin{table}[!htbp]
\centering
\caption{Properties of TDMs for different ``parities'' of eigenfunctions}
\begin{tabular}{|p{0.15\textwidth}<{\centering}|p{0.15\textwidth}<{\centering}|p{0.15\textwidth}<{\centering}|}%
\hline
 $u^s_m(k,r)$ & $u^s_n(k,r)$ & $D^s_m(k)$\\
\hline
 $u^s_m(-k,r)=u^s_m(k,-r)$ &  $u^s_n(-k,r)=u^s_n(k,-r)$ &  $D^s_{mn}(-k)=-D^s_{mn}(k)$\\
\hline
 $u^s_m(-k,r)=u^s_m(k,-r)$ &  $u^s_n(-k,r)=-u^s_n(k,-r)$ &  $D^s_{mn}(-k)=D^s_{mn}(k)$\\
\hline
 $u^s_m(-k,r)=-u^s_m(k,-r)$ &  $u^s_n(-k,r)=u^s_n(k,-r)$ &  $D^s_{mn}(-k)=D^s_{mn}(k)$\\
\hline
 $u^s_m(-k,r)=-u^s_m(k,-r)$ &  $u^s_n(-k,r)=-u^s_n(k,-r)$ &  $D^s_{mn}(-k)=-D^s_{mn}(k)$\\
\hline
\bottomrule
\end{tabular}
\end{table}

From the analysis of Section II, after the ``smooth procedure'', the newly derived eigenfunction $u_{m}^{s}(k, r)$ satisfies the strict periodic boundary condition $u_{m}^{s}(k, r)=e^{i2Gr}u_{m}^{s}(k+2G, r)$ and  $u_{m}^{s}(k, r)=u_{m}^{s}(k, r+R)$. One still would like to know what are the symmetry properties of the TDMs for systems with inversion symmetry. The results are summarized in Table  \uppercase\expandafter{\romannumeral 1} while the derivation is given in the Appendix.   As shown in details in the Appendix,
When the periodic functions $u_{m}^{s}$ and $u_{n}^{s}$ have the same parity, the TDM  between band $m$ and  $n$ is an odd function with respect to k. When they  have  opposite parity, the TDMs between them  is an even function with respect to k.  In a three-band model, there are many pathways to generate excitations. Take the MgO example, one possibility is to choose a path $band  2\rightarrow 3\rightarrow  4$. The corresponding macroscopic polarization is given by
\begin{equation}
\begin{split}
&P(t)\sim P(k_0,t)+P(-k_0,t)\\
&=D_{24}(k_0+A(t))D_{43}(k_0+A(t))D_{32}(k_0+A(t))E(t)^2\\
&+D_{24}(-k_0+A(t))D_{43}(-k_0+A(t))D_{32}(-k_0+A(t))E(t)^2\\
&+c.c.
\end{split}
\end{equation}
Here $E(t)$ and $A(t)$ are electric field and vector potential, respectively. If the driven laser is a long pulse, $E(t+T/2)=-E(t)$ and $A(t+T/2)=-A(t)$, where $T$ is the optical cycle of the laser. Thus, we can get
\begin{equation}
\begin{split}
&P(t+T/2)\sim P(k_0,t+T/2)+P(-k_0,t+T/2)\\
&=D_{24}(k_0-A(t))D_{43}(k_0-A(t))D_{32}(k_0-A(t))E(t)^2\\
&+D_{24}(-k_0-A(t))D_{43}(-k_0-A(t))D_{32}(-k_0-A(t))E(t)^2\\
&+c.c.
\end{split}
\end{equation}
By comparing Eq. (7) and (8) and using the properties listed in TABLE \uppercase\expandafter{\romannumeral 1}, [ the parities of the wavefunctions for these three bands are shown in Fig. 5(b)] we can find that $P(t)=-P(t+T/2)$. The odd parity of macroscopic polarization, similar to the case for an atomic target, guarantees that no even-order harmonics in the spectra.  If the parities of TDMs have not accounted for, odd symmetric macroscopic polarization is not present. In other words, because of the parities of TDMs, coupling of multiple bands cannot generate even harmonics if the system has inversion symmetry.

\begin{figure}[!htbp]
    \includegraphics[width=3.5in,angle=0]{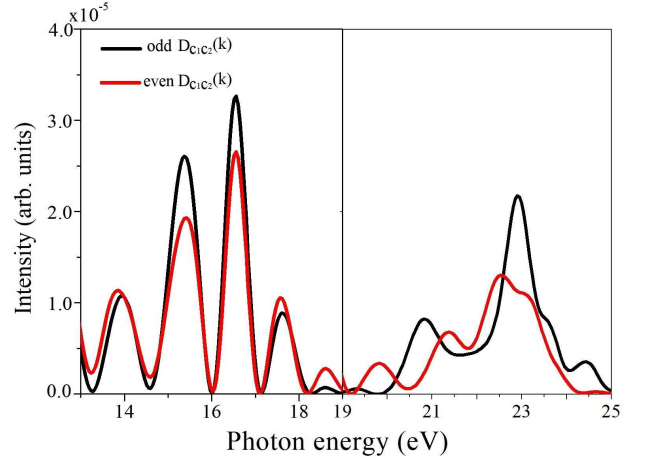}
    \caption{Black line is the same as fig. 3(c): harmonic spectrum calculated using correct dipole moments (D$_{c_1c_2}$(k) is odd, D$_{vc_2}$(k) and D$_{vc_1}$(k) are even functions of k). Red line: The same as the black line except that D$_{c_1c_2}$(k) is changed to an even function of k artificially. The subscripts $v, c_1, c_2$ represent the valence band, first conduction band and second conduction band, respectively. The intensities of the spectra in the right side of the vertical solid line were multiplied by a factor of 4.}\label{Fig_4}
\end{figure}

Using the more accurate band structure and dipole moments, in particular, the newly dipole phases constructed in the present "smooth-periodic" gauge, we can improve the simulation reported in \cite{Shambhu_OL} where the dipole moments have been set to be a constant. Comparison of CEP-dependent spectra between experimental data and our simulation is presented in Fig. 3(a) and (b). The laser parameters used in the simulation can be found in the caption. Many features in the experimental data are reproduced in the present  simulation. (1) In Fig.3(a), the slope of photon energy versus the CEP has been reproduced in Fig. 3(b). (2) Both experiment and simulation show two plateaus, in the same photon energy region. (3) In Fig. 3(b), our simulation indicates a minimum around 13 eV, which is consistent with experiment  even though it is near the low energy end of the experimental data, thus it could also be due to detector efficiency in the energy region. Note that  this minimum also appears in Fig. 2 at the 15th order harmonic. Comparing with the simulations reported in \cite{Shambhu_OL}, we have witnessed significant improvement in the present simulation.

 Figs. 3(c-f) compare the HHG  spectra at two CEPs,   0.5 $\pi$ and 0.0, between experiment  and the present simulation. It is clear that the sine-like pulse (CEP=0.5 $\pi$) would generate  sharper discrete harmonics, while a cosine-like pulse would produce relatively flatter ones. Such results are in agreement with the measurements.

In this article, we are concerned with how harmonic spectra are affected if the parity and periodicity of transition dipole moments are not correctly accounted for. In many prior calculations, approximations were made in which the k-dependent dipole moments were taken to be its absolute values. This means that it is an even function with respect to k. In the three-band model for MgO, using the method presented here, D$c_1c_2$(k), which is the coupling between the two conduction bands, is an odd function. If we arbitrarily change it to an even function, how the HHG spectra would be altered? Fig. 4 shows the original spectra copied from Fig. 3(c) (in black lines) and compare it to HHG spectra (in red lines) if D$c_1c_2$(k) is changed to an even function. For the first plateau harmonics, no significant changes occur since the first plateau is due to recombination of electrons from the first conduction band to the valence band. For the harmonics in the 2nd plateau, we can see the change as the harmonic peaks are shifted. Based on Eqs. (7), (8) and the analysis in \cite{Shambhu_OL}, the peaks in the secondary plateau are  given by: $\omega=(2n-1-\theta /\pi)\omega_0$, where $\omega_0$ is the frequency of the driving pulse and the phase $\theta$ is the phase difference between the two pathways to reach the conduction band 2.   By changing  the parity of D$c_1c_2$(k)   artificially  from odd to even,   the peaks in the 2nd plateau will be given by:  $\omega=(2n-\theta /\pi)\omega_0$. This can explain the shift of the peaks in the secondary plateau shown in Fig. 4. Thus if the parity of the dipole moment has wrong parity, the generated harmonic spectra will be quite different.

We conclude this section by illustrating the difference between gas phase and solids, as shown in Fig. 5.  In the gaseous medium with inversion symmetry, a triangular system will never be formed because of the parity of the wavefunction. For crystal with inversion symmetry where energy level is expanded into a band, a triangle system can be formed. While, the parity of the TDMs will prevent the generation of even-order harmonics.

\begin{figure}[!htbp]
    \includegraphics[width=3.5in,angle=0]{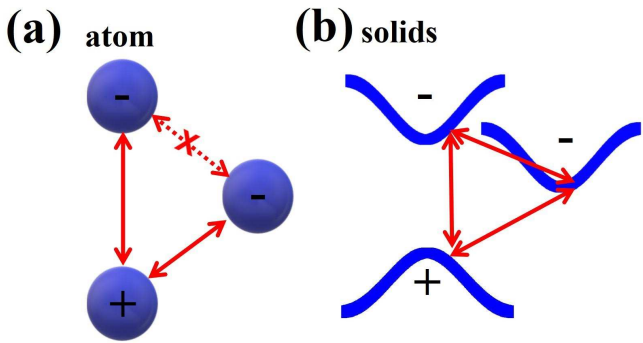}
    \caption{Illustration of transition paths of electrons for gases    and solids   with inversion symmetry.  For the gas  phase, a triangular system will never be formed because the transition between states with same parity is forbidden. While, for solid case where the energy levels are extended into bands, transition is between bands at these k points away from $\Gamma$ are not forbidden. The symmetry properties of the TDMs will ensure the absence of even order optical signal. }\label{Fig_5}
\end{figure}

\section{System with broken symmetry}
For system with broken symmetry, the Zak's phase can be any value. Thus the method above for system with inversion symmetry is not valid anymore. Note that in the periodic gauge the Berry connection is periodic $ D_{m m}(k)=D_{m m}(k+G)$, which means it can be expanded as
\begin{equation}
D_{m m}(k)=\mathrm{g}_{m}(k)+\sigma_{m}
\end{equation}
where $g_m(k)=\sum_{n=1}^{+\infty} f_{1}(n) \cos (n L k)+f_{2}(n) \sin (n L k)$ is the ''AC" component and $\sigma_{m}$ is a constant which can be regarded as the "DC" component.  The DC component will lead to divergence of the introduced additional phase $\Theta_m(k)=\int_{k_0}^{k} d \kappa D_{m m}(\kappa)+\varphi_{m}\left(k_{0}\right)$. We do not  need to care about the AC part, because this component would not influence the periodicity, the continuity and the final observable physical quantities \cite{HCDu}. If the ``smooth procedure'' is carried out for the first Brilloin zone, the DC-induced non-periodic phase of the TDMs between band $m$ and $n$ is $(\sigma_{n}-\sigma_{m})k$.
\begin{figure}[!htbp]
    \includegraphics[width=3.5in,angle=0]{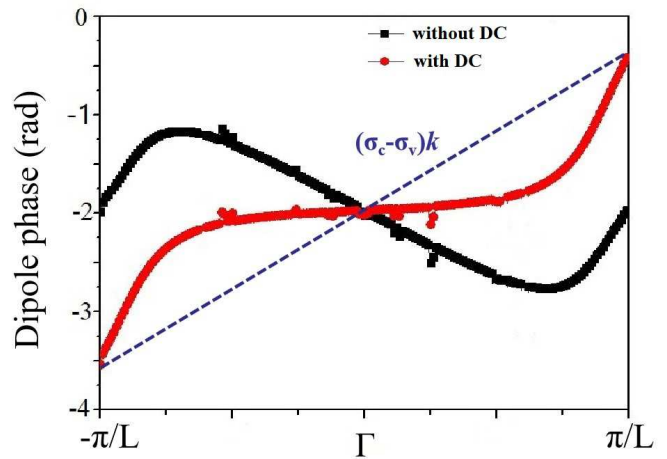}
    \caption{Red line is the k-dependent dipole phase generated by ``smooth procedure''. Such phase is not periodic because of the DC term in the additionally introduced phase. The black line is the k-dependent dipole phase after the DC component is taken away as introduced in this article. }\label{Fig_6}
\end{figure}

Here we take the direction $\Gamma-A$ of wurtzite ZnO as an example. In Fig. 6, the red line is the k-dependent dipole phase for the first Brillouin zone obtained from VASP using a ``smooth procedure''. As stated above, the dipole phase cannot be ensured to be periodic  because of the DC component. However, it is easy to get the slope by  $(\sigma_{n}-\sigma_{m})=(\alpha_{mn}(\pi/L)-\alpha_{mn}(-\pi/L))/G$. Here, $\alpha_{mn}(\pm\pi/L)$ are the phase of TDMs which are read from the data generated by ``smooth procedure'' shown by the red line in Fig. 6.   We can then get the periodic dipole phase by subtracting  the DC part off, $D_{mn}^p(k)=D_{mn}^s(k)e^{i(\sigma_{m}-\sigma_{n})k}$ which is shown by the black line in Fig. 6 for ZnO. At the same time, the Berry connection is changed from zero to $D_{mm}^p(k)-D_{nn}^p(k)=\sigma_{m}-\sigma_{n}$. With that, all the elements in the SBEs model are periodic and continuous with respect to k. This means that  the equation of motion for carriers can be solved in a finite k space.  In Fig. 7, the calculated HHG spectrum (blue line)  by solving two-band SBEs model using TDMs obtained by our ``smooth-periodic'' procedure is compared   to the experimental data (green line). We also present the spectrum (red line) by solving two-band SBEs including only dipole phase obtained from tight-binding model. Even though tight-binding model can
approximatively reproduce the orientation-dependent feature of HHG spectra, usually it is too primitive to produce the relative strength between odd and even order harmonics. Using accurate TDMs obtained from  \emph{ab-initio} software with the help of our ``smooth-periodic'' procedure, the experimental spectra of ZnO first reported in S. Ghimire et al. \cite{shambhu_nature-phys2011} has finally been satisfactorily reproduced theoretically.

\begin{figure}[!htbp]
    \includegraphics[width=3.5in,angle=0]{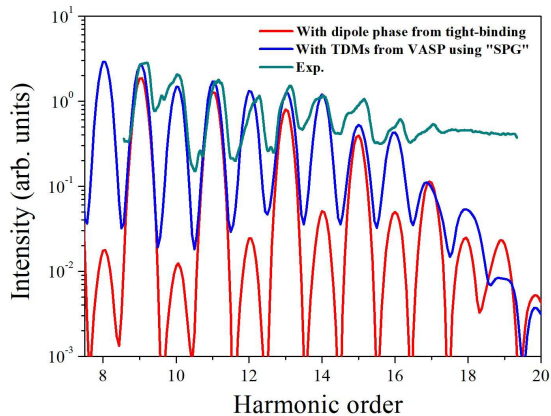}
    \caption{HHG spectra from ZnO. Green line is the experimental data, blue line is calculated by two-band SBEs with elements obtained from \emph{ab-initio} software with the help of the present ``smooth-periodic'' procedure. The red line is from calculations where the dipole phases are calculated from the tight-binding model. The red line and experimental data are copied from our previous work\cite{jiang_prl}. To present clear comparison, the spectra are shifted vertically.  }\label{Fig_7}
\end{figure}

\section{Conclusion remarks}
Although \emph{ab-initio} software has been widely used to investigate electronic properties nowadays, the random phase generated in the algorithms prevents its application to calculate    non-adiabatic dynamics, especially when the external field is a strong laser. In this article, we first show that the commonly used ``smooth procedure'' cannot ensure the periodicity of the wavefunction. Second, we provide two different methods to overcome this defect for systems with and without inversion symmetry. Because our approaches ensure  continuity and periodicity of all the elements used in the equations of motion, the gauge resulting by the transformation of our methods can be referred to as a ``smooth-periodic'' gauge to distinguish it from   ''periodic" gauge used by Resta \cite{Resta}. Based on this gauge,   HHG spectra from solids with and without inversion symmetry are revisited. It is emphasized that   symmetry properties are the key factors to ensure the absence of even order harmonics for systems with inversion symmetry. With the accurate TMDs with dipole phase and Berry connection, the HHG spectrum from ZnO with broken symmetry is also improved greatly. The TMDs introduced in this work is fundamental to all applications relating to optical signals from solids, such as laser waveform control, band/dipole reconstruction and detecting dynamic information.

\section*{Acknowledgment}
This work was supported by the NSF of Jiangsu Province (Grant No. BK20170032) and NSF of China ( 11704187, 11974112, 11975012). CDL was supported in part by the Chemical Sciences, Geosciences, and Biosciences Division, Office of Basic Energy Sciences, Office of Science, U.S. Department of Energy, under Grant No. DE-FG02-86ER13491. SJ also thanks for the support by the Project funded by China Postdoctoral Science Foundation No. 2019TQ0098. SJ thanks for the  fruitful discussion with Prof. Chengcheng Liu from Beijing Institute of Technology and Dr. Prasoon Saurabh from East China Normal University. 

\section*{Appendix: Derivation of the parities of transition dipole moments}

Both $u_{m}^{s}(k, r)$ and $u_{m}^{kp}(k, r)$  satisfy the k$\cdot$p equation,
\begin{align}
\left(-\frac{1}{2} \nabla_{r}^{2}+V(r)-i k \cdot \nabla_{r}\right) u_{m, k}^{s(kp)}(r) =\left(E_{m}(k)-\frac{k^{2}}{2}\right) u_{m, k}^{s(kp)}(r).\tag{A1}
\end{align}
where $u_{m}^{kp}(k, r)$ is assumed to satisfy
\begin{align}
  u_{m}^{kp}(-k, r)=u_{m}^{kp*}(k, r)\tag{A2}
\end{align}
When the system has inversion symmetry, 

\begin{align}
u_{m}^{kp}(-k, r)=\pm u_{m}^{kp}(k, -r); \emph{D}_{\mathrm{mm}}^{{kp}}({k})=0\tag{A3}
\end{align}

$u_{m}^{s}(k, r)$ must be related to $u_{m}^{kp}(k, r)$ through a gauge transformation, e.g. 

\begin{align}
u_{m}^{s}(k, r)=u_{m}^{kp}(k, r)e^{i\beta(k)}. \tag{A4}
\end{align}
As stated in the main text, after the ``smooth procedure'', the Berry connection
\begin{align}
D_{mm}^s(k)= i\left\langle u_{m}^s(k,r) | \nabla_k u_{m}^s (k,r)\right\rangle=0. \tag{A5}
\end{align}
By inserting Eq. (A4) into Eq. (A5), we have
\begin{align}
&\emph{D}_{{mm}}^{{s}}({k})=\boldsymbol{i}\left\langle\emph{u}_{{m}}^{{kp}}({k}, {r}) {e}^{{i} \beta({k})} | \nabla_{{k}}\left(\emph{u}_{{m}}^{{kp}}({k}, {r}) {e}^{{i} \beta({k})}\right)\right\rangle \nonumber\\
&=\emph{D}_{{mm}}^ {{kp}}(k) -\nabla_{{k}} \beta({k})=0\tag{A6}
\end{align}
It is easy to prove that $\emph{D}_{\mathrm{mm}}^{{kp}}({k})$ is a real number and an even function with respect to k,
\begin{align}
\emph{D}_{\mathrm{mm}}^{{kp}}(-{k})=\emph{D}_{{mm}}^{{kp}}({k}). \tag{A7}
\end{align}
In order to ensure $D_{mm}^s(k)=0$, $\beta(k)=\emph{odd function}+const..$
Further, if the system has inversion symmetry,      $\beta(k)=const.$.
Thus, $u_{m}^{s}(k, r)$ also has 
\begin{align}
u_{m}^{s}(-k, r)=\pm u_{m}^{s}(k, -r).\tag{A8}
\end{align}
Using Eq. (A8), we can get all the properties listed in Table \uppercase\expandafter{\romannumeral 1}.

\end{document}